\documentclass{ceab}   

\usepackage{epsfig}     
\usepackage{graphicx}   

\usepackage{ceabbib}     
\usepackage[T1]{fontenc}

\def\XMM{\textit{XMM--Newton}}
\def\RX{RX J0440.9+4431}
\def\TreA{{3A 0535+262}}
\def\aut{\textit{La Palombara et al.}}
\def\tit{\textit{The X--ray soft excess in the low--luminosity HMXRBs}}

\begin{document}

\title{The hot--blackbody spectral excess in low--luminosity High--Mass X--Ray Binaries}

\author{N. La Palombara$^{1}$, S. Mereghetti$^1$, L. Sidoli$^1$, A. Tiengo$^{1,2}$, P. Esposito$^1$
\vspace{2mm}\\
\it $^1$INAF - IASF Milano, via Bassini 15, I--20133 Milano, Italy\\ 
\it $^2$IUSS, Piazza della Vittoria 15, I--27100 Pavia, Italy\\
}

\maketitle

\begin{abstract}
We report on the main results obtained thanks to an observation campaign with \XMM\ of four persistent, low--luminosity ($L_{\rm X} \sim 10^{34}$ erg s$^{-1}$) and long--period ($P > 200$ s) Be accreting pulsars. We found that all sources considered here are characterized by a spectral excess that can be described with a blackbody component of high temperature ($kT_{\rm BB} > 1$ keV) and small area ($R_{\rm BB} < 0.5$ km). We show that: 1) this feature is a common property of several low--luminosity X--ray binaries; 2) for most sources the blackbody parameters (radius and temperature) are within a narrow range of values; 3) it can be interpreted as emission from the NS polar caps.
\end{abstract}


\section{Introduction}

Most of the X--ray binary pulsars (XBPs) are high mass X--ray binaries (HMXRBs), in which a neutron star (NS) with magnetic field B $\sim 10^{12}$ G is accreting matter from a high--mass early--type star (either an OB supergiant or a Be star). They can be persistent sources, with luminosities in excess of 10$^{34}$ erg s$^{-1}$, or transient sources, characterized by quiescent phases with emission around 10$^{34}$ erg s$^{-1}$ or less, occasionally interrupted by bright outbursts reaching $L_{\rm X} \sim 10^{36-38}$ erg s$^{-1}$ \citep{Negueruela98,Reig07,Sidoli10}.

In these sources the X--ray spectrum between 0.1 and 10 keV is usually described by a rather flat power law, with a photon index $\Gamma \sim 1$, but several XBPs have shown a marked \textit{`soft'} X--ray excess above the main power--law component \citep[see][for a review]{LaPalombara&Mereghetti06}. This excess is well described by a thermal emission model (either a blackbody, bremsstrahlung, or hot plasma emission) with low temperature ($kT_{\rm SE} <$ 0.5 keV) and large emission area ($R_{\rm SE}$ larger than a few hundreds km). Only in a few cases this low--energy component has shown coherent pulsations, therefore the debate about its origin remains open.

Here we report on the results we obtained with \XMM\ observations of the four \textit{persistent} Be pulsars originally identified by \citet{Reig&Roche99}, i.e. \mbox{RX J0146.9+6121/LS I +61 235} \citep{LaPalombara&Mereghetti06}, \mbox{4U 0352+309/X Persei} \citep{LaPalombara&Mereghetti07}, RX J1037.5-5647/LS 1698 \citep{LaPalombara+09}, and RX J0440.9+4431/\ LS V +44 17 \citep{LaPalombara+12}; their main parameters are reported in Table~\ref{parameters}. These sources have persistently low luminosity ($L_{\rm X} \sim 10^{34-35}$ erg s$^{-1}$) and long pulse period ($P > 200$ s), two properties which suggest that the NS orbits the Be star in a wide and nearly circular orbit, continuously accreting material from the low--density outer regions of the circumstellar envelope. 

\section{The hot--blackbody spectral excess in persistent Be pulsars}

The \XMM\ spectra of the persistent Be pulsars of our sample cannot be described with a single-component model: the fits with a power-law (PL) or a blackbody (BB) model are affected by large residuals, while other models (e.g. thermal bremsstrahlung or disk blackbody) are rejected by the data. On the other hand, a good fit is obtained with a PL+BB model. In all cases the BB component is characterized by a high temperature ($kT_{\rm BB} > 1.5$ keV) and a small emission radius ($R_{\rm BB} < 0.5$ km), and contributes to 30--40 \% of the source flux below 10 keV (Table~\ref{parameters}): therefore this \textit{hot--blackbody} spectral component can be considered as an additional common property of the persistent Be XBPs, which stands beside those previously known.

\begin{table}[htbp]
\caption{Main parameters of the persistent Be X--ray binaries}\label{parameters}
\begin{center}
\begin{tabular}{lcccc} \hline \hline
Source				& X Persei		& LS I +61 235			& LS 1698			& LS V +44 17			\\ \hline
Spectral Type			& 09.5 IIIe		& B0 IIIe			& B0 III-Ve			& B0.2 Ve			\\
Distance (kpc)			& $\simeq$ 1		& $\simeq$ 2.5			& $\simeq$ 5			& $\simeq$ 3.3			\\
Pulse Period (s)		& 839.3			& 1396.1			& 853.4				& 204.98			\\
$L_{\rm X}^{a}$			& 13.4 $\pm$ 0.1	& $1.45^{+0.03}_{-0.02}$	& 1.15 $\pm$ 0.04		& $7.09^{+0.09}_{-0.08}$	\\
$L_{\rm BB}^{b}$		& $52^{+3}_{-1}$	& $3.5^{+0.4}_{-0.3}$		& $4.9^{+0.9}_{-0.8}$		& $28^{+1}_{-2}$		\\
$L_{\rm BB}/L_{\rm X}$ (\%)	& 39$^{+2}_{-1}$	& 24$^{+3}_{-2}$		& 43$^{+8}_{-7}$		& 39 $\pm$ 2			\\
$T_{\rm BB}$ (keV)		& 1.42 $\pm$ 0.03	& 1.11 $\pm$ 0.06		& 1.26$^{+0.16}_{-0.09}$	& 1.34 $\pm$ 0.04		\\
$R_{\rm BB}$ (m)		& 361 $\pm$ 3		& 140 $\pm$ 15			& 128$^{+13}_{-21}$		& 273 $\pm$ 16			\\
$R_{\rm col}$ (m)		& $\sim$ 330		& $\sim$ 230			& $\sim$ 200			& $\sim$ 320			\\ \hline
\end{tabular}
\end{center}
\begin{small}
$^{a}$ in units of 10$^{34}$ erg s$^{-1}$, in the energy range 0.3--10 keV
\\
$^{b}$ in units of 10$^{33}$ erg s$^{-1}$, in the energy range 0.3--10 keV
\end{small}
\end{table}

On the bases of our results and the emission models proposed by \citet{Hickox+04}, it makes sense to attribute the observed \textit{hot--blackbody} spectral excess to the thermal emission from the NS polar caps. To verify this hypothesis, we assume that the sources are in the `accretor' state, with matter accretion onto the NS surface. Adopting the typical NS parameters $M_{\rm NS}$ = 1.4 $M_{\odot}$, $R_{\rm NS}$ = $10^6$ cm, and $B_{\rm NS} = 10^{12}$ G, from the source luminosity $L_{\rm X}$ we estimate the accretion rate $\dot M$ = [$L_{\rm X}R_{\rm NS}$]/[$GM_{\rm NS}$], the magnetic dipole momentum $\mu = B_{\rm NS}R_{\rm NS}^{3}$/2, the magnetospheric radius $R_{\rm m}$ = [$\mu^{4}$/[$2 G M\dot M^{2}$]$^{1/7}$ \citep{Campana+98}, and then the accretion column radius $R_{\rm col} \sim R_{\rm NS}$ ($R_{\rm NS}/R_{\rm m}$)$^{0.5}$ \citep{Hickox+04}, which is an estimate of the expected size of the polar cap. For each of the four persistent Be X--ray binaries we found that the estimated column radius is remarkably similar to the estimated blackbody emitting radius, a result which strongly supports the idea of a polar--cap origin for the observed BB emission (see Table~\ref{parameters}).

\section{Comparison with other HMXRBs}

\begin{figure}[t!]
\centering
\resizebox{\hsize}{!}{\includegraphics[angle=-90,clip=true]{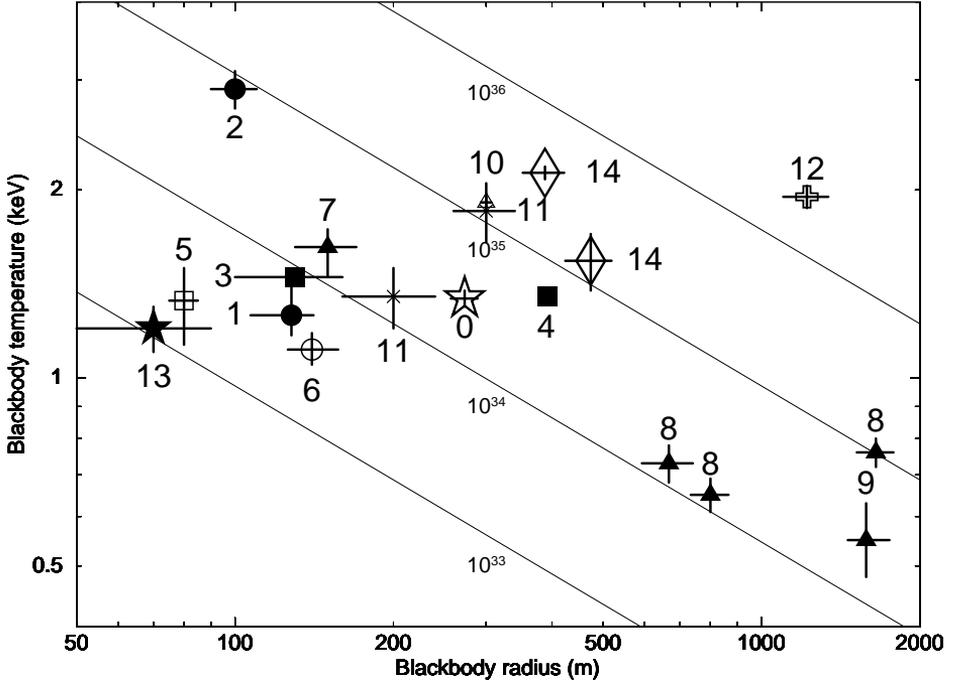}}
\caption{Best--fit values for radius and temperature of the \textit{BB} component observed in the case of \RX\ (0), RX J1037.5-5647 (1,2), 4U 0352+309 (3,4), \TreA\ (5), RX J0146.9+6121 (6), 4U 2206+54 (7,8,9), SAX J2103.5+4545 (10), IGR J11215-5292 (11), IGR J08408-4503 (12), XTE J1739-302 (13), and SXP 1062 (14). The figure shows that, in most cases, the spectral parameters are within a narrow range of values, i.e. $kT_{\rm BB} \sim$ 1--2 keV and $R_{\rm BB} <$ 200 m. The solid lines connect the blackbody parameters corresponding to four different levels of luminosity (in units of erg s$^{-1}$). References: 0 - \citet{LaPalombara+12}; 1- \citet{LaPalombara+09}; 2 - \citet{Reig&Roche99}; 3 - \citet{Coburn+01}; 4 - \citet{LaPalombara&Mereghetti07}; 5 - \citet{MukherjeePaul05}; 6 - \citet{LaPalombara&Mereghetti06}; 7 - \citet{Masetti+04}; 8 - \citet{Torrejon+04}; 9 - \citet{Reig+09}; 10 - \citet{Inam+04}; 11 - \citet{Sidoli+07}; 12 - \citet{Sidoli+09}; 13 - \citet{Bozzo+10}; 14 - \citet{Henault-Brunet+12}.}
\label{BBparameters}
\end{figure}

A spectral feature similar to the \textit{hot BB} of the persistent Be pulsars has been observed also in other low--luminosity ($L_{\rm X} \le 10^{36}$ erg s$^{-1}$) HMXRBs \citep[see][and references therein]{LaPalombara+12}. In Fig.~\ref{BBparameters}, we report the best--fit radius and temperature for the BB component of these sources, together with lines showing four different levels of the blackbody luminosity; each source is displayed with a specific symbol (for some source more than one set of values is shown, corresponding to different observations or flux levels). In most cases, the spectral parameters are within a narrow range of values, i.e. $kT_{\rm BB} \sim$ 1--2 keV and $R_{\rm BB} <$ 200 m. We emphasize that, in all these cases, the estimated total source X--ray luminosity is $\sim 10^{34}$ erg s$^{-1}$, with a 20--40 \% contribution of the blackbody component.

\begin{figure}[t!]
\centering
\resizebox{\hsize}{!}{\includegraphics[angle=-90,clip=true]{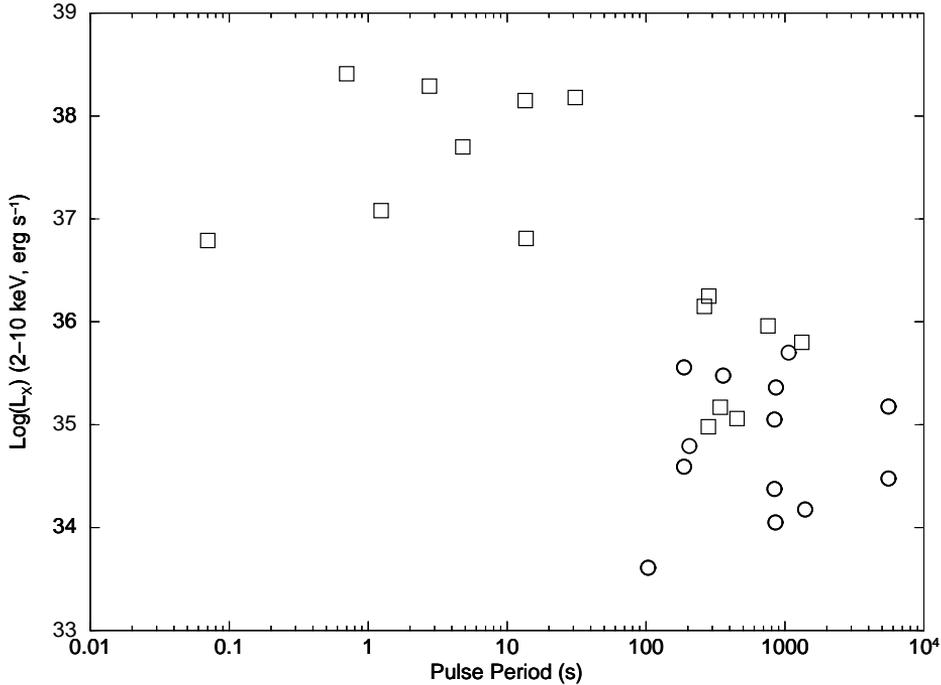}}
\caption{X--ray luminosity (in the 2--10 keV energy range) of the XBPs with a detected thermal excess as a function of the pulse period. The \textit{empty circles} are referred to the \textit{hot BB} pulsars, while the \textit{empty squares} refer to the \textit{soft excess} sources.}
\label{luminosity_period}
\end{figure}

In contrast to this sample of sources, several XBPs are characterized by a \textit{soft} excess, since the fit of this component with a thermal emission model provides low temperatures ($kT_{\rm SE} < 0.5$ keV) and large emitting regions ($R_{\rm SE} > 100$ km). In Fig.~\ref{luminosity_period}, we report the luminosity and pulse period of both types of XBPs: the \textit{soft excess} and the \textit{hot BB} ones are reported as \textit{squares} and \textit{circles}, respectively. On the basis of their distribution in the $P - L_{\rm X}$ diagram, these pulsars are divided into two distinct groups: the sources in the first group are characterized by high luminosity ($L_{\rm X}\ge10^{37}$ erg s$^{-1}$) and short pulse period (\textit{P} $<$ 100 s), and in most cases they are in close binary systems with an accretion disk; those in the second group have low luminosities ($L_{\rm X}\le10^{36}$ erg s$^{-1}$) and long pulse periods (\textit{P} $>$ 100 s), since they have wide orbits and are wind--fed systems. While all the pulsars in the first group have a \textit{soft excess}, both types of pulsars are present in the second group; in this case there is no clear separation between the two types of pulsars, since there is a partial overlap between the \textit{soft excess} and the \textit{hot BB} ones. However, 
the \textit{hot BB} pulsars are the ones that, on average, are characterized by the lowest luminosities and the longest periods. This suggests that the \textit{hot BB} spectral component is a common feature of the low--luminosity and long--period XBPs.

\section*{Acknowledgements} 
This work is based on observations obtained with \XMM, an ESA science mission with instruments and contributions directly funded by ESA Member States and NASA. We acknowledge financial contributions by the Italian Space Agency through ASI/INAF agreements I/009/10/0 (for the data analysis) and I/032/10/0 (for the \XMM\ operations).

\bibliographystyle{ceab}
\bibliography{Proceedings_LaPalombara_HMXRBs}

\end{document}